\documentclass[12pt,preprint]{aastex}

\newcommand{\be}{\begin{equation}}
\newcommand{\ee}{\end{equation}}
\newcommand{\beq}{\begin{eqnarray}}
\newcommand{\eeq}{\end{eqnarray}}

\shorttitle{GOES-SXI loops}
\shortauthors{L\'opez Fuentes, Mandrini \& Klimchuk}

\begin{document}

\title{The temporal evolution of coronal loops observed by GOES-SXI}

\date{ }

\author{M. C. L\'opez Fuentes\altaffilmark{1,2,*}, 
        J. A. Klimchuk\altaffilmark{1},
        C. H. Mandrini\altaffilmark{2,*}} 
\altaffiltext{1}{Naval Research Laboratory, Code 7675, Washington, DC 20375}
\altaffiltext{2}{Instituto de Astronom\'{\i}a y F\'{\i}sica del Espacio,
CONICET-UBA, CC. 67, Suc. 28, 1428 Buenos Aires, Argentina} 
\altaffiltext{*}{Member of the Carrera del Investigador Cient\'{\i}fico,
Consejo Nacional de Investigaciones Cient\'{\i}ficas y T\'ecnicas,
Argentina}

\begin{abstract}
We study the temporal evolution of coronal loops using data from the  
Solar X-ray Imager (SXI) on board the Geosynchronous Operational 
Environmental Satellite 12 (GOES-12). This instrument has the 
advantage of providing continuous 
soft X-ray observations of the solar corona at a high temporal cadence, 
which allows us to follow in detail the full lifetime of each of  
several coronal loops. The observed 
light curves suggest three somewhat distinct evolutionary phases: $rise$,
$main$, and $decay$.  The durations and characteristic 
timescales of these phases [$I/(dI/dt)$, where $I$ is 
the loop intensity] are much longer than a cooling time and indicate that 
the loop-averaged heating rate increases slowly, reaches a maintenance 
level, and then decreases slowly.  It does not turn on or off abruptly.  
This suggests that a single heating 
mechanism operates for the entire lifetime of the loop.
For monolithic (uniform cross section) loops, the loop-averaged heating 
rate is the intrinsic energy release 
rate of the heating mechanism. For loops that are bundles 
of impulsively heated strands, it is an 
indication of the frequency of occurrence of individual heating events, or  
nanoflares.  We show that the timescale of the loop-averaged heating 
rate is proportional to the timescale of the observed intensity variation.  
The constant of proportionality is approximately 1.5 for quasi-steady 
heating in monolithic loops and 1.0 for impulsive 
heating in multi-stranded loops. The ratios of the radiative to 
conductive cooling times in the loops are somewhat less than 1, putting them 
intermediate between the values measured previously for hotter and cooler 
loops.  The new measurements provide further support for the existence of a 
trend suggesting that all loops are heated by the same mechanism, or that 
different mechanisms have fundamental similarities (e.g., are all impulsive or 
are all steady with similar rates of heating).
\end{abstract}

\keywords{Sun: corona -- Sun: magnetic fields -- Sun: flares -- Sun: X-rays}


\section{Introduction}
\label{intro}

 The nature of coronal heating in the Sun and solar-like stars remains one 
of the most important unexplained  
problems of Astrophysics. Presently, it is widely accepted 
that the magnetic field plays a key role in structuring the coronal plasma and 
providing the energy to achieve the observed coronal temperatures. 
However, it remains an open question as to how the coronal heating mechanism 
works.   
Soft X-ray observations of the Sun show a non-uniform emission coming  
mostly from the coronal part of active regions (ARs), particularly 
from so-called coronal loops (Orrall 1981). 
One therefore expects that the analysis of the temporal evolution of observed 
loops would provide some clues for solving the coronal heating problem.

In the simplest approach, one can envision loops 
as monolithic (uniform cross-section) structures in quasi-static equilibrium 
(see e.g., Rosner et al. 1978). In this picture, a slowly varying heat 
source nearly balances the energy losses due to thermal conduction and 
radiation. 
If the temperature, $T$, emission measure, $EM$, loop diameter, $d$, and loop 
length, $L$, are known, one 
can obtain estimates of the timescales involved in conductive and radiative  
cooling. A comparison between cooling times and 
the typical evolutionary timescale for a loop can then be 
used to test whether the quasi-static interpretation is reasonable. 
Following this approach, Porter \& Klimchuk (1995, henceforth PK95) studied a 
set of 47 loops observed by the Soft X-ray Telescope (SXT, 
Tsuneta et al. 1991) 
on board \textit{Yohkoh}. In this study they followed individual loops for only
a single spacecraft orbit, 
$\sim$ 1 hr of daylight, and 
determined evolutionary timescales, 
$I/(dI/dt)$, where $I$ is 
the loop intensity.  These are ``instantaneous" timescales in the sense that 
the intensity and its derivative are averages 
over a short time interval, shorter than the timescale itself.  
Since the computed
timescales turned out to be typically one or two orders of magnitude longer 
than the 
cooling times, the authors concluded that the loops could be in quasi-static 
equilibrium.
Nevertheless, there remained the 
question of whether the loops actually persist for as long as their  
instantaneous evolutionary timescales suggest. 

  Recent studies based on observations from the Extreme Ultraviolet Imaging 
Telescope (SoHO/EIT, Delaboudiniere et al. 1995) and the Transition Region 
and Coronal Explorer (TRACE, Handy et al. 1999) show that most warm loops 
($T \sim$ 10$^{6}$ K) have densities that are not consistent 
with quasi-static equilibrium (Aschwanden et al. 1999, Aschwanden et al. 
2001, Winebarger et al. 2003). One possibility is 
that coronal loops are bundles of  
multiple unresolved strands, each of which is independently heated in
impulsive events, such as nanoflares (Cargill 1994, Klimchuk 2002, 2006, 
Warren et al. 2003, Cargill \& Klimchuk 2004). This agrees with
Parker's conjecture (1988) that 
energy is released impulsively by reconnection at localized magnetic 
discontinuities in the corona. These discontinuities are thought to be 
formed by the tangling of elemental magnetic ropes via photospheric granular 
motions. A major goal in coronal physics is to determine whether loops are 
bundles of strands that are heated by nanoflares or monolithic structures 
that are heated in a quasi-steady fashion. 

Due to the lack of prolonged and continuous observations of individual 
loops, PK95 were not able to study the lifetimes of loops or how loops 
evolve during their lifetime (i.e., how the evolutionary timescale 
changes).  In particular, they could not investigate 
how loops are ``born'' and how they ``die''.  
A complete understanding of coronal heating requires that this be 
explained.   
V. Kashyap and R. Rosner (1994, unpublished results) attempted to
study the birth of soft X-ray loops using SXT images,
but a combination of data gaps and inadequate cadence prevented them
from drawing any definitive conclusions.
Litwin \& Rosner (1993) pointed out 
that if loops turn on suddenly, the heating rate required to produce the 
hot plasma initially far exceeds the heating rate required to maintain 
its temperature 
in the presence of radiative and conductive losses. In other words, 
the heating profile describing how the heating rate varies with time must 
have a sharp spike followed by a lower level steady phase. 
Loops that turn on gradually, on the other hand, would indicate a
heating rate that ramps up slowly to the maintenance level.
Similarly, loops that fade rapidly (on the cooling timescale) indicate
a heating that shuts off abruptly, and loops that fade gradually
indicate a heating rate that diminishes slowly (see e.g., Serio et al.
1991, Jakimiec et al. 1992). 

This view must be modified somewhat if loops are multi-stranded 
structures heated by many small nanoflares.  In that case, a sudden 
loop brightening would mean that a flurry of 
nanoflares commences all at once, while a gradual turn on 
would indicate a steadily growing frequency of nanoflares.
This last possibility could be related to the development of a
self-organized critical system (L\'opez Fuentes et al. 2005; Klimchuk,  
L\'opez Fuentes, \& DeVore 2006).  In the context of such a picture, 
a gradual fading of the loop might indicate that 
criticality has been lost.  Nanoflares would continue to occur, but with 
steadily diminishing frequency.

 In this work, we analyze data from the Solar X-ray Imager (SXI) 
on board the Geosynchronous Operational Environmental Satellite 12 (GOES-12). 
This instrument has the advantage over SXT, used 
by PK95, that it can observe the solar corona continuously.  
Thus, despite having a lower spatial resolution, 
it is ideal for addressing the problems discussed above. Our primary objective 
is to study the long-term evolution of coronal loops and, thereby, to infer 
the time-dependent 
properties of the coronal heating mechanism. We study the complete life 
cycle of loops, including their birth, maintenance, and decay.  We compute 
evolutionary timescales and cooling times, and compare them in order 
to test for quasi-static equilibrium. Due to the particular sensitivity 
of the SXI filters, the observed loops have physical properties ($T$ and $EM$) 
that are intermediate between those of SXT and TRACE loops. 
We discuss how SXI loops fill a gap in a trend suggested by SXT and TRACE 
loops, which seems to support a picture of impulsive heating by nanoflares 
(Klimchuk 2004, 2006). 

  The paper is organized as follows. In Section~\ref{data}, we describe 
the characteristics of SXI and the data used here. 
In Section~\ref{analysis}, we present the analysis of the observed loops:
we obtain light curves for the full loop evolution 
(Section~\ref{evolution}), we compute $T$ and $EM$ 
(Section~\ref{diagnostics}), and we calculate the radiative and 
conductive cooling times for the observed loops 
(Section~\ref{cooling_times}). In 
Section~\ref{heating} we discuss the implications of our results 
for coronal heating and 
derive a relationship  between the observed 
light curves and the evolution of the heating for the quasi-static 
case, or the frequency of nanoflares for the impulsive heating case. 
These relationships
can be used to test the validity of coronal heating models.   
We discuss our results  
and conclude in Section~\ref{discussion}. 

\section{Description of the data}
\label{data}

The SXI instrument was built by NASA's Marshall Space Flight Center 
with funding from the US Air Force 
(Hill et al. 2005, Pizzo et al. 2005). It is a broadband imager 
in the 6-60 \AA~bandpass that produces full-disk solar images with
$\sim$ 1 min cadence. The images consist of $512 \times 512$ 
pixel arrays with 5 arcsec resolution. The FWHM of the telescope point spread 
function is 
$\sim$ 10 arcsec. 
A set of selectable thin-film entrance filters allows plasma temperature 
discrimination:  Open, 3 polyimide (thin, medium and thick) and 3 
beryllium (thin, medium and thick). Open and polyimide filters are sensitive 
to plasmas 
below 2 MK. Periods of orbital eclipse for GOES-12 are both widely spaced and 
brief (less than 2\% of the total time); therefore, it is especially 
well suited for 
continuous tracking of coronal loops.

\begin{figure*}              
   \centering
  \hspace{0cm}
\includegraphics[bb= 13 13 440 440,width=10cm]{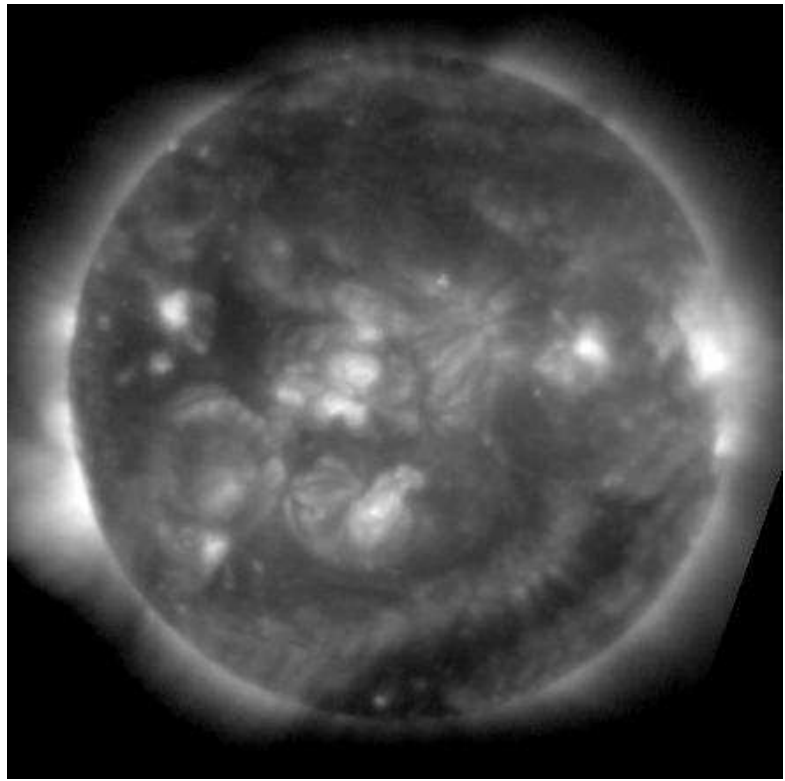}
      \caption{GOES-SXI image of the solar disk obtained with the Open filter
at 00:10 UT on 28 August, 2003. Some of the studied
loops can be seen in this image (see Fig.~\ref{loops}).}
         \label{full_disk}
\end{figure*}

After scanning the SXI database in search of continuous observations of 
non-flaring ARs available in several filters, we selected 
data from 26--28 August, 2003. These observations were obtained 
before the hardware degradation suffered by SXI on 5 November, 2003, which 
affected its operation capabilities and possibly its spectral response 
(Hill et al. 2005).
Figure~\ref{full_disk} shows an image obtained at 00:10 UT 
on 28 August, 2003. Seven ARs were present on the solar disk 
during the selected dates. 
We created a movie of 457 coaligned images that were observed with the 
Open filter at a rate of 6 per hour.  The images were 
rotated to have solar North up and shifted to a common disk center 
position using the SolarSoftware package and
other programs that we developed.  


\section{Loop analysis}
\label{analysis}

\subsection{Loop evolution}
\label{evolution}

\begin{figure*}              
   \centering
  \hspace{0cm}
\includegraphics[bb= 55 200 510 660,width=13cm]{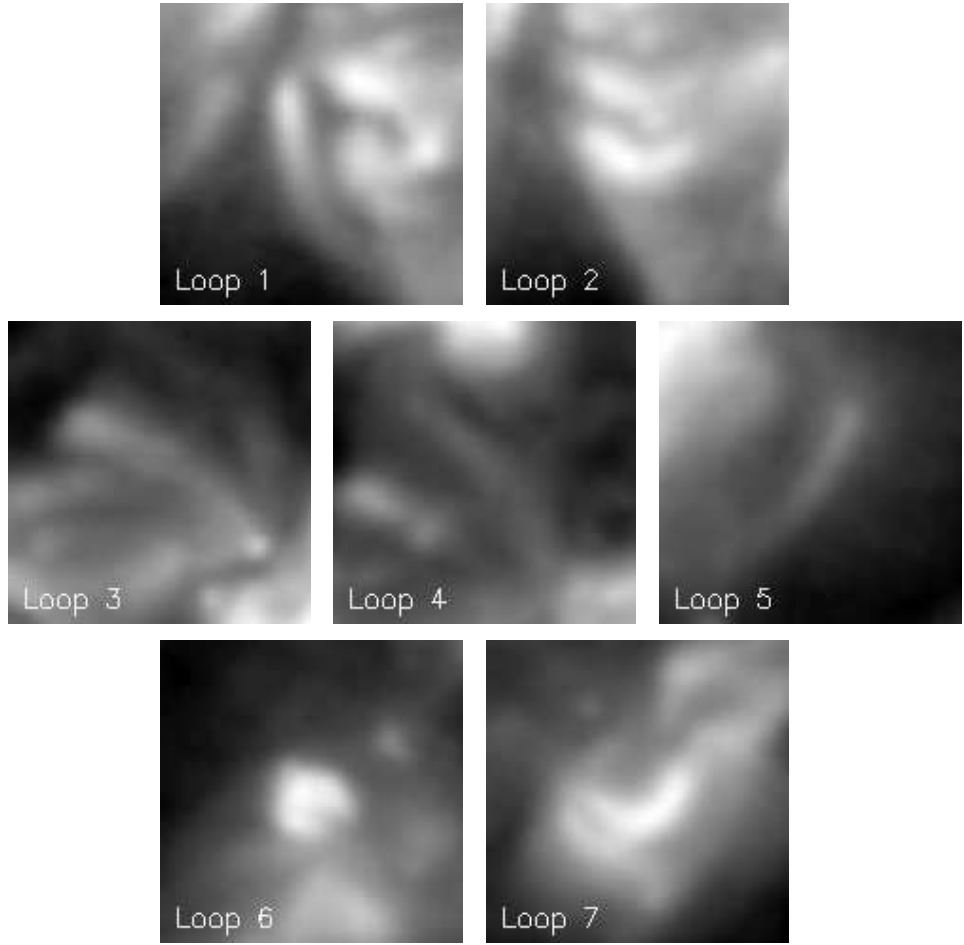}
      \caption{Closeup Open filter images of the SXI loops used in the 
study. On each image the corresponding loop is located 
at the center of the frame. The images correspond to the times of maximum 
loop intensity (during the 
main phase of their evolution, see Section~\ref{evolution}) and display
a square field of view having 100 Mm side.}
         \label{loops}
\end{figure*}

 Seven loops were selected from the movie for detailed study.  They are  
shown in Figure~\ref{loops} at the times of maximum intensity. 
These particular loops were chosen because they are 
relatively well isolated from other overlapping structures
that could contaminate the measurements (for a discussion of the 
necessity of isolated loops, see e.g., Reale \& Ciaravella 2006), 
and because there were no C-class or larger flares reported at the 
location of the loops.

\begin{figure*}              
   \centering
  \hspace{0cm}
\includegraphics[bb= 15 15 230 250 ,width=8cm]{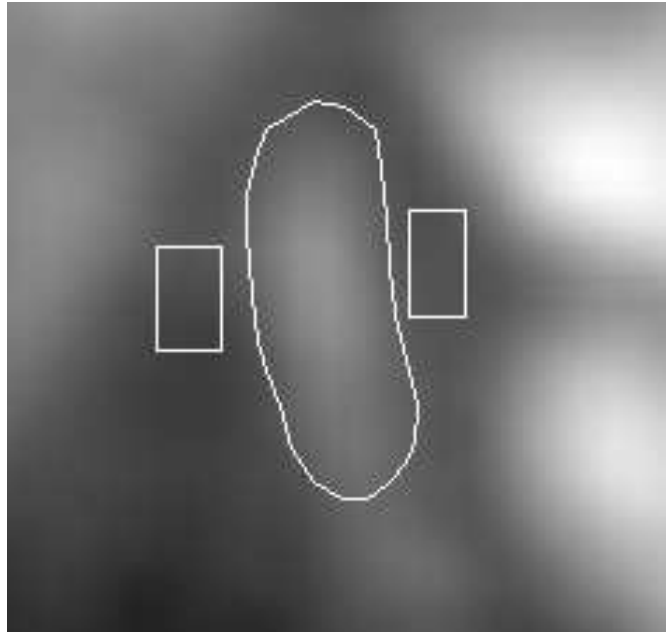}
      \caption{ 
Regions used to determine the loop and background intensities for Loop 1 
(see Figure~\ref{loops}).  The average intensity per pixel from the square 
background regions is subtracted from the loop region to obtain the 
intrinsic loop intensity.}
         \label{measure}
\end{figure*}

  To study the temporal evolution of the loops, we generated light curves 
using the following procedure. We first constructed sets of 
subimages that are corrected for differential 
rotation.  This was necessary because most of the loops were observed for 
roughly half a day, while one could be followed for more than 24 hours.   
Since the corona is optically thin in soft X-rays, not all the emission
along the line of sight comes from the structure we are interested in; thus, 
we substracted a background intensity from each of our loops.   
On a reference subimage of each loop set, 
we subjectively selected three areas, one covered by the loop and two 
by background 
regions on either side (see Fig.~\ref{measure}). We computed the average 
intensity of the pixels in each of these areas.
We then subtracted the average background from the average loop intensity
to obtain the intrinsic intensity of the 
loop. We repeated the process using the same loop and background areas for 
all the images of the set. Extreme care was taken in the subjective selection
of the areas, so they always include only loop emission or only background 
emission for all of the images in the set. In no case did the actual 
loop drift out of the contour defined in the first image. 
Finally, we obtained light curves of the intrinsic intensity for each of the 
selected loops.  

Loops are not uniformly bright, as can be seen in Figure~\ref{loops} and as 
discussed by, for example, Reale \& Ciaravella (2006) and  
Di Giorgio et al. (2003).  We selected 
the brightest part of the loop to make the intrinsic intensity measurements 
(e.g., Fig.~\ref{measure}).  In no case did the bright area appear to 
move during the lifetime of 
the loop. Propagating fronts have been observed in some TRACE loop 
studies, such as Reale et al. (2000), but there is no evidence
of them in the analyzed SXI loops.

\begin{figure*}              
   \centering
  \hspace{0cm}
\includegraphics[bb= 54 84 565 766,width=16cm]{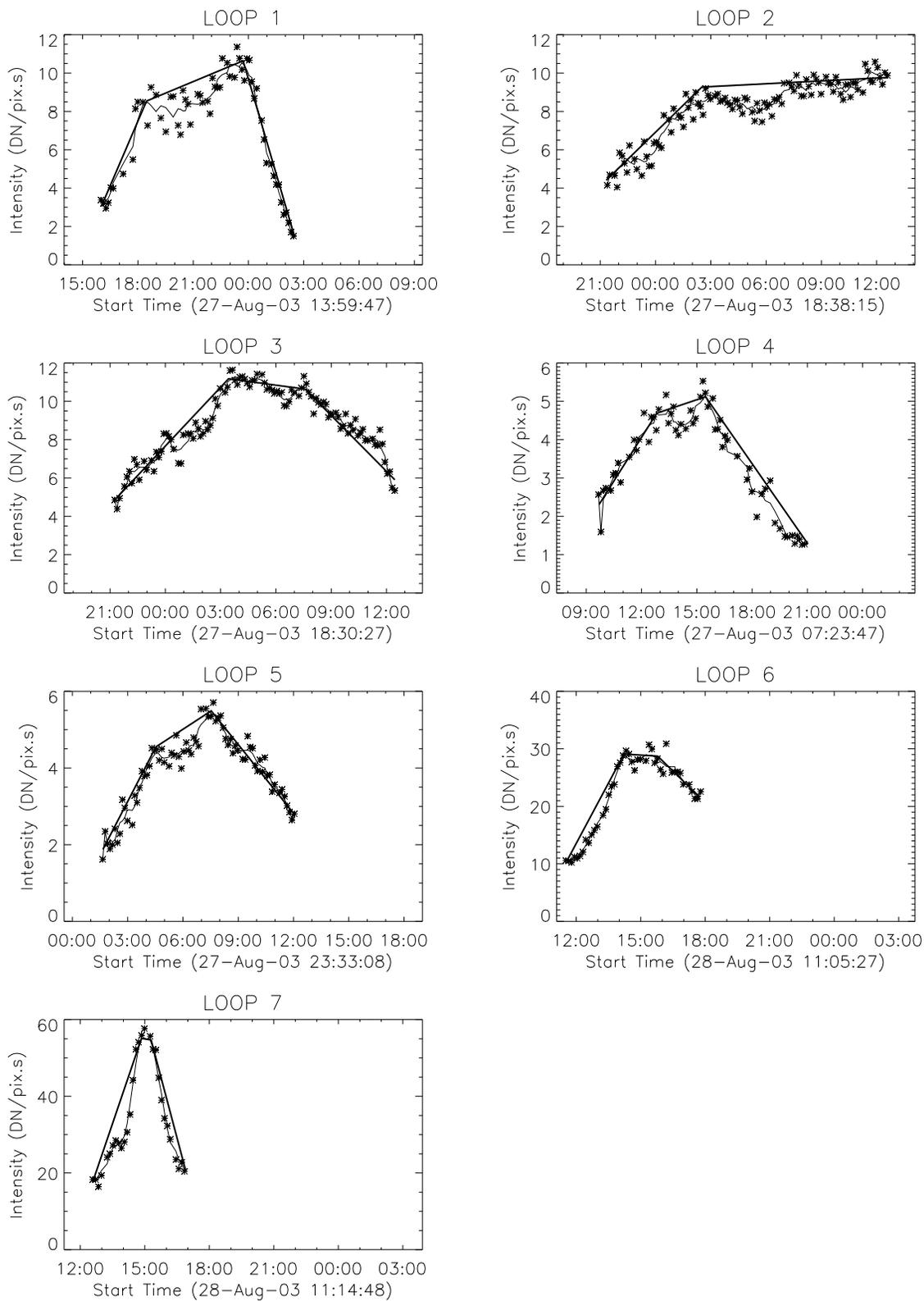}
      \caption{\footnotesize{Lightcurves of all the observed loops constructed from 
the intrinsic (background subtracted) intensities obtained from sets 
of coaligned images. Asterisks correspond to individual
measurements (one per image); the solid line is a
5 point-running average; and the thick straight lines mark the 
subjectively identified rise, main, and decay phases of 
the evolution (see Section~\ref{evolution}).}}
         \label{lightcurves}
\end{figure*}

Figure~\ref{lightcurves} shows the light curves of all seven loops, 
plotted on a similar horizontal scale for easy comparison.  Only part 
of the light curve of loop 2 is shown because it lived much longer than 
the others.  The asterisks 
correspond to individual intensity measurements and the continuous lines 
are 5-point-running averages to highlight the overall evolution. 
It can be seen that the loops go through three rather 
distinct phases, which we define as the $rise$, $main$, 
and $decay$ phases.  We subjectively identify these phases by noting 
the times when there 
is a qualitative change in 
the slope of the smoothed light curve. The phases are marked with 
thick straight lines in the panels of Figure~\ref{lightcurves}.
Table~\ref{properties} lists the durations of the 
phases for all seven loops, together with the loop lifetime, which we define to span from the start of the rise phase to the 
end of the decay phase (i.e., the sum of the rise, main, and decay phase 
durations).  The main phase of loop 7 is not clear, and it is possible 
that this loop evolves from a rise phase directly into a decay phase.  

All of the light curves begin and end at a significant positive value of 
the intensity.  This is somewhat surprising, since it is the intrinsic 
(background-subtracted) intensity that is plotted, and we might expect it 
to be closer to zero and even slightly negative in some cases.  We offer 
two possible explanations.  First, the plasma volume 
that we identify as the loop may have a low level of sustained heating 
that precedes the rise phase and continues beyond the decay phase.  
Second, the emission outside the loop volume may be brighter in the 
immediate vicinity of the loop, including just below it and above it, than 
in the more distant areas to the side that we have chosen to define 
the level of background emission.  

The second possibility raises the issue of the uncertainty in the background 
subtraction and therefore the uncertainty in the intrinsic loop 
intensities we are trying to measure.  We have assumed that the intensities  
to either side of the loop accurately characterize the background along the 
loop line of sight.  One indication of the uncertainty in the background is 
the short-term variability in the side intensities.  
Table~\ref{properties} lists for each loop the root-mean-square 
(rms) intensity variation of the side areas with respect to the
10-point running average.  The values are 
given as a fraction of the intrinsic loop intensity averaged over the main 
phase. We consider these to be an estimate of the uncertainty in the 
intrinsic loop intensity.  The average 
uncertainty for the seven cases is 6\%. The 
table also lists the rms variation of the intrinsic intensity during the main 
phase relative to the main phase average. The mean value for all the loops is 
7\%, and we conclude that most of the measured variability is due to errors 
in the background subtraction. However, since our objective is to study the 
longer-term evolution of loops, the details of the short-term variability 
are not important. 
We note that the background evolves more slowly and with smaller amplitude 
than do the loops themselves.

 Following PK95, we compute evolutionary timescales according to: 
\be
\tau_{evol} = \frac {I_{m}}{(\Delta I/\Delta t)}, 
\ee
\noindent
where $I_{m}$ is the mean intensity 
and $\Delta I/\Delta t$ is the mean intensity derivative for a given time 
interval, $\Delta t$.
PK95 were only able to consider a single short ($< 1$ hr) interval for each 
loop, but we derive characteristic 
timescales for the full durations of the rise, main, and decay phases.  They 
are based on the straight line segments drawn subjectively to match the 
actual light curve, as shown in thick lines in Figure~\ref{lightcurves}.

The values of $\tau_{evol}$ for the three evolutionary phases are listed 
in Table~\ref{properties}.  They are positive or negative depending on 
whether the intensity is increasing or decreasing, respectively. 
The durations of the phases are positive, by definition. The timescales and 
durations tend to be similar to each other for both the rise and decay
phases. For the main phase, on the other hand, 
the timescales are 1--2 orders of magnitude longer
than the durations.  This is because the slope of the light curve 
is very small during the main phase. 
Many of the timescales measured by PK95 are 
very long (hundreds of hours), and it is likely that those loops were observed 
during their main phase. Note that the intensity can be either 
increasing or decreasing during the main phase.  This may  
indicate that the energy input to the loop can be 
either very slowly increasing or very slowly decreasing during this time.
However, we cannot rule out the possibility that imperfect background 
subtraction contributes to this result. 

  Litwin \& Rosner (1993) argued that if loops are monolithic and 
turn-on suddenly, the onset time should be comparable to the sound-transit 
time (see also Fisher, Canfield, \& McClymont 1984).
For typical coronal loops the sound-transit time is of the order  
a few tens of seconds, which is much shorter than the typical 
rise phase timescales of our loops.  We conclude that the 
loop-averaged heating must turn on slowly (though it could turn on quickly 
within the individual strands of a multi-stranded loop).  As we will show 
in Section~\ref{cooling_times}, 
the cooling times of our loops are much shorter than the 
decay phase timescales, so we also conclude that the 
loop-averaged heating turns off slowly.
Considering the full evolution of the loops, it would seem that 
the heating must be of a similar nature in all three phases, including 
the main phase.

\subsection{Loop temperatures and densities}
\label{diagnostics}

In order to compute cooling times, for eventual comparison with 
the evolutionary timescales, we must first determine the temperatures and 
emission measures of the loops.  
As is customary with broadband instruments, $T$ and $EM$ 
are estimated using the filter ratio technique, which 
takes advantage of the fact that 
observations made with different wavelength bandpasses have different 
temperature sensitivities.  The observed intensity is given by 
\be
\label{int}
I = \int n^{2} S(T)~dV = EM~S(T)
\ee
\noindent
for an isothermal plasma, where $dV$ is the differential emitting volume, 
$n$ is the electron density, and $S(T)$ is the instrument response  
(for SXI see http://www.sec.noaa.gov/sxi/).
For two different filters ($a$ and $b$) with response functions 
$S_{a}(T)$ and $S_{b}(T)$, the ratio of the intensities 
depends only on temperature:
\be
\label{ratio}
R(T) = \frac{S_{a}(T)}{S_{b}(T)} = \frac{I_{a}}{I_{b}}. 
\ee
\noindent
Thus, from an observed filter ratio 
$R(T)$ one can obtain an estimate of the temperature under the 
assumption that the plasma is isothermal. For a multi-thermal plasma, 
the inferred temperature will be a complicated weighted average, with 
greater weighting where the filters are more sensitive.

\begin{figure*}              
   \centering
  \hspace{0cm}
\includegraphics[bb= 79 360 555 720,width=15cm]{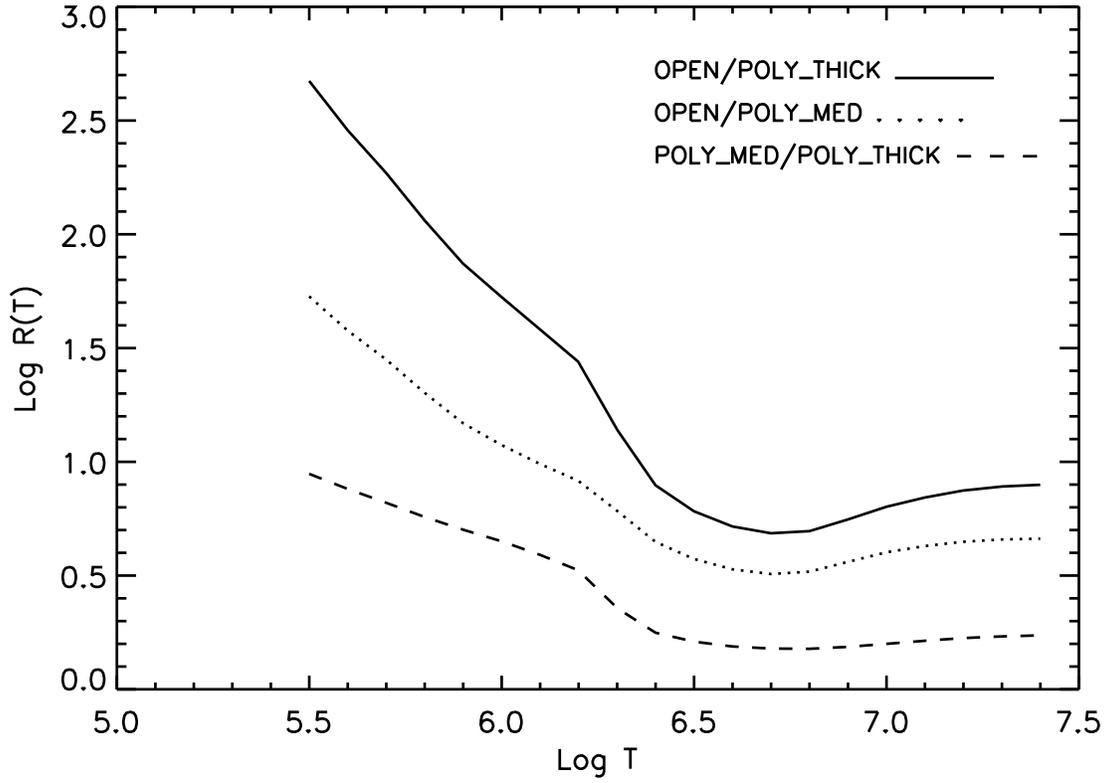}
      \caption{Filter ratio versus temperature for three different filter
combinations. Open/thick-polyimide is the most suitable combination 
for our purposes since it has the steepest dependence on $T$
in the range of maximum sensitivity of the filters ($T<2$ MK).}
         \label{ratios}
\end{figure*}

 Since there are several possible filter combinations for 
$T$ and $EM$ diagnostics, we must decide which one is
the most suitable for our observations.
Figure~\ref{ratios} shows plots of $R$ versus $T$ for 
three different combinations: Open/thick-polyimide, 
Open/medium-polyimide and thick-polyimide/medium-polyimide. 
The Open/thick-polyimide 
combination has the steepest dependence on $T$ in the range ($\leq$ 2 MK)
of maximum sensitivity of the filters. 
Therefore, we select that filter combination and compute 
the ratio of intrinsic loop intensities to obtain $T$ using
Equation~\ref{ratio}. Once $T$ is known, $EM$ is obtained from 
Equation~\ref{int} using either of the two filters.
Thick-polyimide images are taken approximately once 
every 7 hours. Since most of the studied loops live longer than
that, we can compute at least one and sometimes two $T$ values
for each loop.  The first value is from the  
main phase, and the second value, if available, is 
from near the end of the rise phase or near the beginning of the 
decay phase. The two $T$ values turn out 
to be similar within error bars, and we average them to obtain 
a single temperature for the loop.  The individual temperature values   
are themselves averages in that they are computed from spatially averaged loop 
intensities. The filter intensities in Equation~\ref{ratio} 
are obtained following the procedure described in Section~\ref{evolution}. 

 Once we have estimates of $T$ and $EM$, we can 
obtain the electron number density,
\be
\label{density}
n = \left( \frac{EM}{fd} \right) ^{1/2},
\ee
and the plasma pressure,
\be
P = 2nkT,
\ee
\noindent
where $f$ is the filling factor (the fraction of the observed volume
occupied by emitting plasma), $d$ is the loop diameter, and $k$ is 
the Boltzmann constant. Here, we assume $f = 1$, so our density is a 
lower limit.
We determine $d$ following the procedure described in Klimchuk et al.
(1992), PK95 and Klimchuk (2000), which is based on the calculation of the 
standard deviation (i.e., second moment) 
of the intensity profile (after background subtraction) 
in a direction perpendicular to the loop axis.  It can be demonstrated that
the standard deviation is $\frac{1}{4}$ of the loop 
diameter for well-resolved observations of uniformly filled loops having 
circular cross-sections.  Actual loop observations suggest that the 
assumptions of a circular cross-section and uniform density (on 
resolvable scales) are reasonable (see Klimchuk 2000, 
L\'opez Fuentes et al. 2006). Thus, for each loop we use the image of 
maximum loop brightness to compute the standard 
deviation of the intensity profile at every pixel position along the 
loop axis. We then obtain an average diameter for the loop that is four 
times the average of the standard deviations. Our measurements 
show that these values do not vary significantly along the loops
(see also, Klimchuk 2000, L\'opez Fuentes et al. 2006).

The inferred temperatures and densities of the observed loops
are shown in Table~\ref{properties}. The temperatures  
range between 1.2 and 2.3 MK, and the densities are of order  
$10^9~$cm$^{-3}$. These loops are therefore
cooler and more dense than most {\it Yohkoh}/SXT loops (PK95) and 
slightly hotter 
and less dense than most TRACE loops (Aschwanden et al. 1999, 
Aschwanden et al. 2001, Winebarger et al. 2003). 

  Our determinations of loop physical parameters have uncertainties 
that depend on the uncertainties in the 
intrinsic intensity measurements. 
The nominal 
error of SXI intensities, which is based on photon statistics and 
detector properties, is discussed 
by Pizzo et al. (2005). The uncertainty associated with 
$N_{phot}$ detected photons is $\sqrt{2N_{phot}}$. 
We propagate these uncertainties through the $T$,
$EM$, and $n$ computations, taking full account of the fact that 
the $T$ and $EM$ uncertainties are not independent, but rather are 
correlated through the filter ratio $R(T)$.
See Klimchuk \& Gary (1995) for 
a complete analysis of the errors associated with the filter ratio 
technique.  We follow a similar 
approach here. The resulting uncertainties 
in $T$, $EM$, and $n$ are typically 10\% or less.  
We have not formally accounted for the 
uncertainties associated with the background subtraction, which 
depend primarily on our subjective choice of the background areas, 
but we have verified that the intrinsic loop intensities do not change 
significantly when 
the background areas are shifted by a few pixels in each direction.

\subsection{Cooling timescales}
\label{cooling_times}

  The energy equation for coronal loops in quasi-static equilibrium 
describes a balance of thermal conduction, optically thin radiation, and 
an unknown mechanism of coronal heating, and is given by:
\be
\label{energy_eq}
Q + \kappa_{o} \frac{d}{ds} \left( T^{5/2} \frac{dT}{ds} \right)
             -  n^{2} \Lambda(T) = 0,
\ee 
\noindent
where $Q$ is the volumetric heating rate, $s$ is the curvilinear coordinate 
along the loop, and we have neglected variations in the 
loop cross-sectional area. 
The second and third terms account for 
conductive and radiative losses, respectively. 
For those terms, $\kappa_{o} T^{5/2}$ is the thermal 
conductivity (with $\kappa_{o} =$ 10$^{-6}$ erg s$^{-1}$ cm$^{-1}$ 
K$^{-7/2}$, Spitzer 1962), and
$\Lambda(T)$ is the optically thin radiation loss function.   

  An estimation of the cooling times can be 
obtained by dividing the energy density of the gas by the energy loss 
rate, $\mathbb{L}$, associated with  
the corresponding process:
\be
\label{tau}
\tau \approx \frac{\frac{3}{2}P}{\mathbb{L}}. 
\ee
\noindent
If we are only interested in the gross properties of the loop, the conduction 
term can be approximated as:
\be
\mathbb{L}_{cond} = \kappa_{o} \frac{d}{ds} \left( T^{5/2} \frac{dT}{ds} 
   \right) \approx \frac{2}{7} \kappa_{o} \frac{T^{7/2}}{L^{2}}.
\ee
\noindent
where $L$ is the loop half-length and $T$ is the average loop temperature 
under the assumption that $T$ is almost constant along most of the loop 
(see PK95 and references therein). Loop half-lengths have been estimated by 
visually identifying the locations of the footpoints in the straightened 
loop image used in the diameter measurement procedure. They are listed in Table~\ref{properties}.  Substituting the 
above expression into Equation~\ref{tau} gives the conductive cooling time:
\be
\tau_{cond} \approx \frac{21}{4}~\frac{k}{\kappa_{o}} 
                    ~\frac{PL^{2}}{T^{7/2}}.
\ee
  For the radiative cooling time, the radiative loss function $\Lambda(T)$ 
can be expressed as a power of $T$ (Vesecky et al. 1979, Kano 
\& Tsuneta 1995):
\be
\Lambda(T) = \Lambda_{o} T^{b},
\ee
\noindent
where $\Lambda_{o}$ and $b$ are constants. Combining this with 
Equation~\ref{tau} and the corresponding term in Equation~\ref{energy_eq} 
($\mathbb{L}_{rad} = n^{2} \Lambda(T)$), we obtain:
\be
\tau_{rad} \approx \frac{6k}{\Lambda_{o}}~\frac{T^{2-b}}{P}. 
\ee
\noindent
The constants $\Lambda_{o}$ and $b$ for the temperature range of the 
loops studied here are:
$\Lambda_{o} = 1.9 \times 10^{-22}$ and $b = 0$ for $T \leq 1.51$ MK,
and $\Lambda_{o} = 3.53 \times 10^{-13}$ and $b = -1.5$ for $T > 1.51$ MK
(PK95).

  Finally, the net cooling time due to 
radiation and conduction acting together is given by:
\be
\label{cool_time}
\tau_{cool} = \left(\tau_{cond}^{-1} + 
                \tau_{rad}^{-1}\right)^{-1}.
\ee
\noindent
The computed $\tau_{cool}$ values are given in Table~\ref{properties}. 
The uncertainties, obtained by propagating 
the errors in the physical parameters, are of order 5\%.
Note that these cooling times correspond to the main phase, or very close 
to it. 
Using a simple analytical model, Serio et al. (1991) obtained 
a slightly different expression for the cooling time that is
approximately 0.7 times $\tau_{cool}$ given in Equation~\ref{cool_time}.
This alternate form does not change 
the conclusions below.

We see from Table~\ref{properties} that the evolutionary timescales 
of the three different phases are much longer than the cooling times: 
4-15 times longer for the rise phase; 24-174 times longer for the 
main phase; and 2-13 times longer for the decay phase.  In comparison, the 
evolutionary timescales of the loops studied by PK95 (although not stated
in the paper) are between 2 and 9300 
times longer than the cooling times, with a median value of 23.  This suggests 
that most of the PK95 loops were observed during the main phase, as already 
noted.  
The table also shows that the durations of the evolutionary phases, including 
the main phase, are longer than the cooling times. These results 
indicate that the loops could be in quasi-static equilibrium.

   Detailed solution of the hydrodynamic equations reveals 
that thermal conduction is a slightly stronger cooling mechanism than 
radiation in 
static equilibrium loops and therefore that $\tau_{rad}/\tau_{cond} \gtrsim 1$ 
(e.g., Vesecky et al. 1979).   The last row 
in Table~\ref{properties} gives the ratios computed for our loops.  They 
are all less than, but not much less than, unity.  This indicates that 
radiation is somewhat stronger than thermal conduction in these loops and, 
therefore, that they are not in quasi-static equilibrium.  This can be 
reconciled with the result that the evolutionary timescales are much longer 
than the cooling times if we consider the loops to be bundles of unresolved 
impulsively-heated strands, as discussed below. 

Figure~\ref{rad_cond} is a plot of the cooling time ratio,  
$\tau_{rad}/\tau_{cond}$, versus temperature.  The seven SXI loops studied 
here are shown as diamonds, while the crosses come from measurements 
of {\it Yohkoh}/SXT loops at higher temperatures and TRACE loops at lower 
temperatures (Klimchuk 2006). The SXI values are intermediate between the 
SXT and TRACE values and help to define what appears to be a continuous 
diagonal band of points.  There do {\it not} seem to be two physically 
distinct classes of loops.

It is interesting to speculate that the continuous distribution of loop 
properties indicated by Figure~\ref{rad_cond} implies that all loops are 
heated by the same mechanism, or that different mechanisms have fundamental 
similarities (e.g., are all impulsive or are all steady with similar rates of 
heating).  If impulsive heating and steady heating were both common, we would 
expect Figure~\ref{rad_cond} to have a diagonal band of points from the 
impulsively heated loops (Klimchuk 2006) and a horizontal band of points at 
$\tau_{rad}/\tau_{cond} \gtrsim 1$ from the steadily heated loops.  Furthermore, 
if there existed both strong and weak mechanisms of steady heating, 
we would expect a cluster of points at high temperature and a cluster of 
points at lower temperature, with few loops in between.  
Additional study of the frequency of different temperature loops is  
needed before anything definitive can be claimed.


\section{Possible heating scenarios}
\label{heating}

\begin{figure*}              
   \centering
  \hspace{-1.cm}
\includegraphics[bb= 15 15 495 430 ,width=15cm]{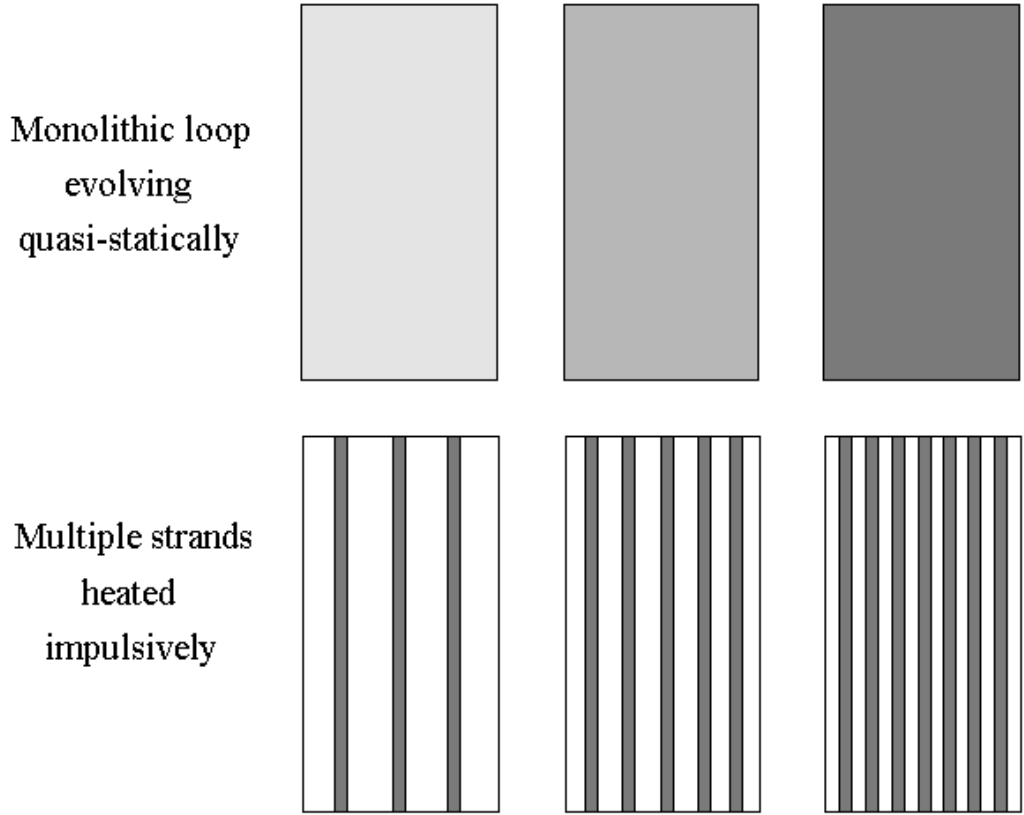}
      \caption{Sketch of the rise phase of a loop  according to the two 
different heating scenarios discussed in Section~\ref{heating}. 
The upper panel corresponds to a monolithic loop that evolves 
quasi-statically as the heat source slowly and uniformly
provides energy. 
The lower panel shows a loop comprised of many unresolved strands
that are heated impulsively. The intensity increases with the 
number of heated strands (i.e., the nanoflare occurrence rate).}
         \label{drawing}
\end{figure*}

  As we have already mentioned in the Introduction, there are two 
basic scenarios for the heating of slowly evolving loops.  In one case, 
loops are monolithic structures (or they are 
multi-stranded with all of the strands being identical), and the heat 
source changes slowly so that the plasma adjusts in a 
quasi-static fashion.  In the other case, loops are bundles of 
impulsively heated strands, and the frequency of nanoflare events 
changes slowly.  
In this way the loop will appear to evolve gradually, 
even though the individual strands do not. A schematic representation 
of these two possibilities is shown in  
Figure~\ref{drawing} for the rise phase of a loop. In the upper panel, 
the intensity of
the loop increases due to a slowly varying heat source that provides
energy uniformly. The level of loop brightness, indicated by the 
hatching density, increases with time. 
In the lower panel, the number of elemental strands that 
are heated, i.e. the number of nanoflares, increases with time with each 
strand having the same brightness.  
It is possible to quantitatively 
relate the change in the loop intensity to either the change in the volumetric 
heating rate (monolithic loop) or the change in the nanoflare frequency 
(multi-stranded loop), as we now discuss.

\subsection{Monolithic loops in quasi-static equilibrium}
\label{mono-quasi}

  In quasi-static equilibrium, the three terms in 
the energy equation (Equation~\ref{energy_eq}) are roughly equal 
(Vesecky et al. 1979):
\be
\label{rates}
Q \sim \frac{2}{7} \kappa_{o} \frac{T^{7/2}}{L^{2}}
         \sim n^{2} \Lambda_{o} T^{b}.
\ee

  From Equation~\ref{int}, we know that the loop intensity is related to the 
instrument response function $S(T)$ by:
\be
\label{int_vs_resp}
I \propto n^{2} S(T).
\ee
\noindent
For the temperature range of the studied loops and the Open filter, the 
response function can be written as:
\be
\label{response}
S(T) = S_{o} T^{a},
\ee
\noindent
where $S_o$ and $a$ are constants, such that $a = 0.93$ for $T \le 1.51$ MK,
and $a = 0.54$ for $T > 1.51$ MK.
Combining the expression for $S(T)$ with Equations~\ref{rates} 
and~\ref{int_vs_resp}, we obtain  the following relationship 
between the volumetric heating rate and the loop intensity:
\be
Q \propto I^{\phi},
\ee
\noindent
where the exponent $\phi$ is given by:
\be
\phi = \frac{1}{1+\frac{2}{7}(a-b)} = \left\{ 
       \begin{array}{ll}
        0.79 & 1MK \le T \le 1.51 MK \\
        0.63 & 1.51MK < T \le 2.5 MK. 
        \end{array} \right.
\ee
\noindent
Therefore, the intensity profile (i.e., light curve) is a direct indication 
of the volumetric heating rate profile. The heating rate timescale, 
$\tau_{Q}=Q/(dQ/dt)$, and the intensity timescale, $\tau_{I}=I/(dI/dt)$, 
are then related in the following way:
\be
\label{steady}
\tau_{Q} = \phi^{-1} \tau_{I}.
\ee
The heating rate timescale is approximately 1.5 times longer than the 
intensity timescale.

It is important to bear in mind that the above 
analysis is valid only when the heating changes slowly enough that 
the physical conditions in the loop are close to static equilibrium.  
That is what is meant by quasi-static equilibrium.  As a rule of thumb, 
quasi-static equilibrium applies whenever the heating rate timescale is 
several times longer than both the cooling timescale and the end-to-end sound 
transit timescale.  These are the characteristic times required to achieve 
energy balance and force balance, respectively.  As shown in 
Table~\ref{properties}, these conditions are met in our loops.

\subsection{Multiple strands heated by nanoflares}
\label{multiple-nano}

 Let us now consider loops that are bundles 
of impulsively heated elementary strands. We assume that the 
shape of the nanoflare energy distribution does not change during the 
lifetime of the loop, so that the ratio of large nanoflares to small 
nanoflares is the same at all times.  We also assume that all 
strands cool completely before being reheated.  Under those conditions, the 
average temperature of the heated strands will also not change. The 
intensity of the loop bundle is then 
a direct indication of the nanoflare occurrence rate:
\be
I \propto \frac{dN}{dt},
\ee

\noindent
where $dN$ is the number of nanoflares being produced at interval $dt$.
Since the global heating rate of the loop (integrating 
over the strands) is proportional the rate of nanoflare occurrence, the 
timescale for the global heating rate will be 
equal to the intensity timescale:
\be
\label{impulsive}
\tau_{Q} = \tau_{I}.
\ee
Note that this relationship will not be accurate at the very start of the 
rise phase and very end of the decay phase, when the light curve will 
reflect the evolution of individual strands.

   Equations~\ref{steady} and~\ref{impulsive} impose observational 
constrains on theories of coronal heating.  Any complete theory must 
be able to explain the entire observed light curves of loops, 
including the different 
timescales during the rise, main, and decay phases.


\section{Discussion and conclusions}
\label{discussion}

  The main aim of this work has been to investigate the long-term evolution of 
non-flaring coronal loops,
from their initial brightening to their full decay. Soft X-ray observations 
taken by SXI, though
having poorer spatial resolution than SXT and TRACE, have the advantage of 
extended and effectively continuous temporal coverage that allows us to 
undertake such a study without data gaps. Through a detailed analysis 
of seven loops observed from 26-28 August, 2003, we were able to verify 
and greatly broaden 
previous conclusions (PK95) that were based on short-term SXT observations.

  We found that the temporal evolution of all our studied loops can be 
separated into three 
phases: $rise$, $main$ and $decay$. The durations of the individual phases 
range from half an hour to as long as 7 hours. The total loop lifetimes 
range between about 4 hours and 1 day. The timescales that characterize 
the intensity variations range between roughly 3 and 16 hours for both 
the rise and 
decay phases and are 1-2 orders of magnitude longer for the main phase.  
An important result is that all of these timescales are considerably 
longer than the cooling time and sound-transit time 
(see Table~\ref{properties}). 

We can draw the following conclusions from these results. First, the 
loop-averaged heating rate must increase slowly, reach a nearly constant 
level, and then decrease slowly.  It is not the case that the 
heating turns on suddenly, which, as pointed out by Litwin \& Rosner (1993), 
would imply that the heating has 
two dramatically different phases: an intense initial phase followed by 
a moderate maintenance phase.  The fact that the heating changes 
slowly during all three evolutionary phases suggests that a single heating 
mechanism operates for the entire lifetime of the loop.  
 
Considering that loops are formed by unresolved magnetic strands, 
the loop-averaged heating rate may be very different from the heating rate 
within individual strands. If 
the strands are all identical, so that the loop is a monolithic structure, 
then the local and global heating rates are the same.  However, recent 
observational and theoretical evidence suggests that the strands are 
heated impulsively and are very different from each other at any instant 
in time (e.g., Klimchuk 2006 and references cited therein). In that case, 
the entire 
loop bundle may appear to evolve slowly even if the individual strands are 
evolving rapidly.  The loop-averaged heating rate then reflects 
the frequency with which individual heating events (nanoflares) are occurring, 
rather than the intrinsic variation of the heating rate within each strand.  
We have shown that the timescale of the loop-averaged heating rate 
is proportional to the timescale of the observed intensity variation under 
both scenarios.  The constant of proportionality is approximately 1.5 for 
quasi-steady heating in monolithic loops and 1.0 for impulsive 
heating in multi-stranded loops. 

\begin{figure*}              
   \centering
  \hspace{-1.cm}
\includegraphics[bb= 90 360 550 700,width=15cm]{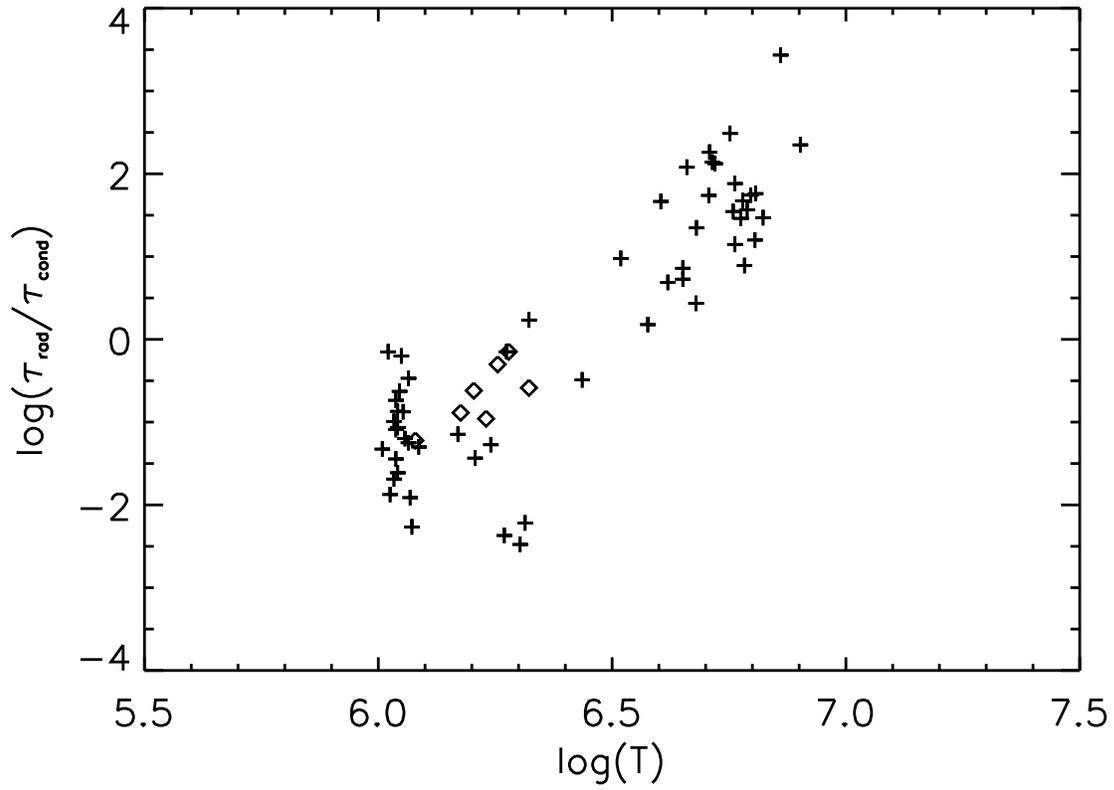}
      \caption{Ratio of radiative to conductive cooling times versus
loop temperature. Pluses correspond to {\it Yohkoh}/SXT and 
TRACE loops, while diamonds correspond to the SXI loops studied here.  
Adapted from Klimchuk (2006).}
         \label{rad_cond}
\end{figure*}

Previous studies have shown that hot SXT loops ($T > $ 2$\times$ 10$^{6}$ K) 
are under dense  compared to what is expected from quasi-static equilibrium 
solutions (PK95). On the other hand, warm TRACE (and SoHO/EIT) loops 
(T $\sim$ 10$^{6}$ K) are 
over dense (Aschwanden et al. 1999, Aschwanden et al. 2001, 
Winebarger et al. 2003). This is reflected in the ratio of the 
radiative to conductive cooling times shown in Figure~\ref{rad_cond},
where SXI loops are identified with diamonds while the cluster of plus 
signs at higher temperatures correspond to SXT
loops and those at lower ones to TRACE.  
Hot loops have a large ratio, indicating that thermal conduction dominates 
at low densities, while warm loops have small ratios, indicating that 
radiation dominates at high densities. Both results can be 
understood in terms of strands that are cooling after having been 
impulsively heated to high temperatures; SXT sees only the hot strands, 
and TRACE and EIT see only the warm strands.

SXI loops also fit naturally into this nonequilibrium picture.  They are 
of intermediate temperature, and there is a correspondingly smaller 
imbalance between conduction and radiation.  Radiation is 
modestly stronger than conduction in SXI loops, whereas conduction would be  
modestly stronger than radiation if they were in static equilibrium.  The 
SXI loops begin to fill a gap between the SXT and TRACE points in 
Figure~\ref{rad_cond}.  They provide further evidence that all loops are 
physically related and may be heated by a common mechanism 
(e.g., Klimchuk 2006).


If loops are indeed bundles of impulsively heated strands, then we should 
observe SXT, SXI, and TRACE/EIT loops together at the same physical location.  
There will be a temporal shift, since it takes time for the plasma to cool 
from SXT to SXI to TRACE/EIT temperatures, 
but, depending on the duration of the ``storm'' of nanoflares, 
we may expect some overlap of the light curves detected by the different  
instruments (e.g., Winebarger \& Warren 2005; Ugarte-Urra et al. 2006). 
Unfortunately, SXT and TRACE observations are not available 
for the loops in our study.  The Yohkoh spacecraft was no longer 
operating, and TRACE was observing different active regions elsewhere 
on the solar disk.  We do have a limited number of EIT observations in 
essentially the same 171 A band of TRACE.  The cadence was very 
slow, so there are only one or two EIT images for each SXI 
loop.  The visibility of the loops in the EIT images can be summarized as 
follows.  Loops 1 and 4 can be seen during the decay phase, but 
not at the beginning of the main phase.  Loop 5 can be seen at the end 
of the main phase.  Loops 2, 3, and 6 can be seen during the rise 
phase, but only as pairs of bright areas corresponding to the SXI footpoints. 
Loop 2 is also visible during the main phase, still as footpoint emission, 
but loop 3 is not.  Loop 7 is short lived and does 
not coincide with any EIT images.  Despite the better resolution, the EIT 
versions of the loops do not show significant differences in structuring 
or morphology with respect to their SXI counterparts, except of course 
in the cases where only footpoint emission is visible. 
Whether these multi-instrument observations 
are consistent with bundles of impulsively heated strands can only be answered 
with detailed modeling, which is beyond the scope of the present study.

Among the theoretical evidence of nanoflare heating is the demonstration 
that the secondary instability produces explosive energy release when 
the misalignment angle between adjacent magnetic flux tubes reaches a 
critical value (Dahlburg, Klimchuk, \& Antiochos 2005).  It is well 
known that the photospheric field is concentrated into elemental flux 
tubes of kG strength (e.g., S\'anchez Almeida \& 
Lites 2000).  A single SXI loop must contain many hundreds to thousands of 
these tubes.  Parker (1988) pointed out that the footpoints of the tubes are 
randomly displaced by turbulent convection, causing the field to become 
tangled in the corona.  Magnetic energy is apparently released when the
misalignment angle of the intertwined tubes reaches the secondary 
instability threshold.

This basic picture has many similarities to cellular automata models of 
self-organized critical (SOC) systems. We have recently begun to explore 
whether the observed SXI light curves can be explained by a simple 
SOC model (L\'opez Fuentes et al. 2005; Klimchuk, L\'opez Fuentes, 
\& DeVore 2006).  
Our initial results are encouraging. We imagine that the coronal field is 
originally untangled. Photospheric convection shuffles the footpoints and 
progressively increases the tangling. In the early stages, relatively few 
pairs of adjacent tubes reach the secondary instability threshold, and 
the SXI  intensity is relatively faint. As time proceeds, the emission 
brightens as the number of tube pairs reaching the threshold steadily 
increases. 
This is the rise phase. Eventually, a self-organized critical state is 
achieved in which the stressing of the field by convection is balanced by 
the destressing of the field by impulsive energy releases. This is the main 
phase, when the intensity 
is roughly constant. Finally, something causes the effectiveness of the 
footpoint shuffling to diminish, and the loop enters a decay phase when the 
critical state can no longer be maintained. The cause of this loss of 
effectiveness is not entirely clear. One possibility is that the concentration 
of kG tubes decreases as the tubes diffuse across the photosphere. 
Footpoint shuffling is efficient at tangling the field only when the random 
walk step size is longer than the characteristic footpoint separation. 
When this condition is not met, a random step will not necessarily wrap a 
flux tube around its neighbor and there is a reduced likelihood that the 
level of tangling will increase.  

We are currently investigating the dependence of the predicted light curve 
on the various parameters of the model and plan to report the results 
in an upcoming paper.
The observed SXI light curves described here provide important 
constraints, not only for this model, but all models of coronal heating.


\acknowledgements
We acknowledge the GOES/SXI team, in particular Vic Pizzo and
James McTiernan for their helpful advise on the handling and 
interpretation of SXI data. We thank our referee Fabio Reale
for his enriching suggestions and comments. C.H.M. thanks the 
members of the Solar-Terrestrial Relationships 
branch at NRL for their hospitality during her stay there. 
This work has been funded by NASA and the 
Office of Naval Research. 



\clearpage
\begin{table}
\caption{Properties of the studied loops. All time values are given in 
hours. Loop and background rms are relative to the averaged loop intrinsic 
intensity during the main phase.}
\label{properties}
\vspace{0.5cm}
$\begin{array}{lrrrrrrr}
 & $Loop 1$ & $Loop 2$ & $Loop 3$ & $Loop 4$ & $Loop 5$ & $Loop 6$ & $Loop 7$ \\
\hline 
$Date$             &  8/27 &  8/27 &  8/27 & 8/27 & 8/28 & 8/28 & 8/28 \\
$Start time (UT)$  & 16:00 & 21:20 & 21:10 & 9:40 & 1:40 & 11:30 & 12:30 \\
$Temperature (MK)$ & 1.5 & 1.6 & 1.7 & 1.8 & 1.2 & 1.9 & 2.1 \\
$Density $(10^{9} cm^{-3}) & 0.9 & 1.3 & 1.1 & 0.9 & 1.1 & 1.2 & 1.9 \\
$Loop length $(10^{9} cm) & 11 & 5.8 & 12 & 7.2 & 9.1 & 5.4 & 6.5 \\
$Lifetime $       & 10.9 & > 27.4 & 15.2 & 11.6 & 10.9 & 6.4 & 4.2 \\
$Rise duration $  &  3.2 &    5.2  &  6.2  &  3.6   &  3.6 &   3.2 & 2.2 \\
$Rise timescale $  & 3.6 &    8.5  &  7.8  &  5.7   &  4.7 &   3.4 & 2.2 \\
$Main duration $  & 4.7 &     ...  &  4.2  &  3.1   &  2.9 &   1.4 & 0.5     \\
$Main timescale $ & 21.3 &    ...  & -83.0 &  132.2 & 16.7 & -48.1 & -31.4 \\
$Decay duration $ &  3.0 &    ...  &  4.9  &  4.9  &  4.4 &   1.8 & 1.5  \\
$Decay timescale $ & -2.0 &    ... & -8.5  & -4.5  & -7.6 &  -9.7 & -1.7 \\
$Loop rms$       & 0.08 & 0.04 & 0.03 & 0.08 & 0.06 & 0.06 & 0.13 \\
$Background rms$ & 0.05 & 0.04 & 0.04 & 0.08 & 0.04 & 0.02 & 0.12 \\
$Cooling timescale $ & 0.89 & 0.57 & 0.85 & 0.76 & 0.64 & 0.57 & 0.53 \\
\tau_{rad}/\tau_{cond} & 0.33 & 0.24 & 0.11 & 0.50 & 0.06 & 0.71 & 0.26 \\
$Sound-transit time$ & 0.013 & 0.013 & 0.011 & 0.014 & 0.013 & 0.02 & 0.028 \\
\hline
\end{array}$
\end{table}


\clearpage

\begin{thebibliography}{}

\bibitem[Aschwanden et al.(1999)]{1999ApJ...515..842A} 
Aschwanden, M.~J., Newmark, J.~S., Delaboudini{\` e}re, J.-P., Neupert, 
W.~M., Klimchuk, J.~A., Gary, G.~A., Portier-Fozzani, F., \& Zucker, A.\ 
1999, \apj, 515, 842 

\bibitem[Aschwanden et al.(2001)]{2001ApJ...550.1036A} 
Aschwanden, M.~J., Schrijver, C.~J., \& Alexander, D.\ 2001, \apj, 550, 1036 

\bibitem[Cargill(1994)]{1994ApJ...422..381C} 
Cargill, P.~J.\ 1994, \apj, 422, 381 

\bibitem[Cargill \& Klimchuk(2004)]{2004ApJ...605..911C} 
Cargill, P.~J., \& Klimchuk, J.~A.\ 2004, \apj, 605, 911 

\bibitem[Dahlburg et al.(2005)]{2005ApJ...622.1191D} 
Dahlburg, R.~B., Klimchuk, J.~A., \& Antiochos, S.~K.\ 2005, \apj, 622, 1191 

\bibitem[Delaboudiniere et al.(1995)]{1995SoPh..162..291D} 
Delaboudiniere, J.-P., et al.\ 1995, \solphys, 162, 291

\bibitem[Di Giorgio et al.(2003)]{2003A&A...406..323D} Di Giorgio, S., 
Reale, F., \& Peres, G.\ 2003, \aap, 406, 323 

\bibitem[]{05}
Fisher, G. H., Canfield, R. C., \& McClymont, A. N.  1984,  
\apjl, 281, L79 

\bibitem[Handy et al.(1999)]{1999SoPh..187..229H} 
Handy, B.~N., et al.\ 1999, \solphys, 187, 229 

\bibitem[Hill et al.(2005)]{2005SoPh..226..255H} 
Hill, S.~M., et al.\ 2005, \solphys, 226, 255

\bibitem[Jakimiec et al.(1992)]{1992A&A...253..269J} Jakimiec, J., 
Sylwester, B., Sylwester, J., Serio, S., Peres, G., \& Reale, F.\ 1992, 
\aap, 253, 269 
 
\bibitem[Kano \& Tsuneta(1995)]{1995ApJ...454..934K} 
Kano, R., \& Tsuneta, S.\ 1995, \apj, 454, 934 

\bibitem[Klimchuk(2000)]{2000SoPh..193...53K} 
Klimchuk, J.~A.\ 2000, \solphys, 193, 53

\bibitem[Klimchuk(2002)]{2002ITPConf}
Klimchuk, J.~A.\ 2002, in G. Fisher and D. Longcope (eds.), ITP Conf. on Solar 
Magnetism and Related Astrophysics, U. California Santa Barbara 
(http://online.kitp.ucsb.edu/online/solar\_c02/klimchuk/).

\bibitem[Klimchuk(2004)]{01} 
Klimchuk, J. A.\ 2004, in Proceedings of the SOHO 15 Workshop:  Coronal 
Heating (ESA SP-575), ed. J. Ireland \& R. Walsh (Noordwijk:  ESA/ESTEC), 2 

\bibitem[Klimchuk(2006)]{02} 
Klimchuk, J. A.\ 2006, \solphys, 234, 41

\bibitem[Klimchuk \& Gary(1995)]{1995ApJ...448..925K} 
Klimchuk, J.~A., \& Gary, D.~E.\ 1995, \apj, 448, 925 

\bibitem[Klimchuk et al.(1992)]{1992PASJ...44L.181K} 
Klimchuk, J.~A., Lemen, J.~R., Feldman, U., Tsuneta, S., \& Uchida, Y.
\ 1992, \pasj, 44, L181 

\bibitem[]{03} 
Klimchuk, J.~A., L\'opez Fuentes, M.~C., \& DeVore, C.~R. \ 2006, in Proceedings of SOHO-17:  Ten Years of SOHO and Beyond (ESA SP-617), ed. H. Lacoste (Noordwijk:  ESA/ESTEC)

\bibitem[Litwin \& Rosner(1993)]{1993ApJ...412..375L} 
Litwin, C., \& Rosner, R.\ 1993, \apj, 412, 375 
 
\bibitem[L\'opez Fuentes et al.(2005)] {} 
L\'opez Fuentes, M. C., Klimchuk, J. A. \& D\'emoulin, P.\ 2006, \apj, 639, 459

\bibitem[]{04} L\'opez Fuentes, M. C., Klimchuk, J. A., \& Mandrini, C. H.
2005, Eos. Trans. AGU, 86(18), Jt. Assem. Suppl., Abstract SP14A-06

\bibitem[Orrall(1981)]{1981sars.work.....O} 
Orrall, F.~Q.\ 1981, Solar Active Regions: A monograph from Skylab Solar 
Workshop III  

\bibitem[Parker(1988)]{1988ApJ...330..474P} 
Parker, E.~N.\ 1988, \apj, 330, 474 

\bibitem[Pizzo et al.(2005)]{2005SoPh..226..283P} 
Pizzo, V.~J., et al.\ 2005, \solphys, 226, 283 

\bibitem[Porter \& Klimchuk(1995)]{1995ApJ...454..499P} 
Porter, L.~J.~\& Klimchuk, J.~A.\ 1995, \apj, 454, 499 

\bibitem[Reale et al.(2000)]{2000ApJ...535..412R} Reale, F., Peres, G., 
Serio, S., DeLuca, E.~E., \& Golub, L.\ 2000, \apj, 535, 412 

\bibitem[Reale \& Ciaravella(2006)]{2006A&A...449.1177R} Reale, F., \& 
Ciaravella, A.\ 2006, \aap, 449, 1177 

\bibitem[Rosner et al.(1978)]{1978ApJ...220..643R} 
Rosner, R., Tucker, W.~H., \& Vaiana, G.~S.\ 1978, \apj, 220, 643 

\bibitem[S{\' a}nchez Almeida \& Lites(2000)]{2000ApJ...532.1215S} 
S{\'a}nchez Almeida, J., \& Lites, B.~W.\ 2000, \apj, 532, 1215 

\bibitem[Serio et al.(1991)]{1991A&A...241..197S} Serio, S., Reale, F., 
Jakimiec, J., Sylwester, B., \& Sylwester, J.\ 1991, \aap, 241, 197 

\bibitem[Spitzer(1962)]{1962Spitzer} 
Spitzer, L. \ 1962, Physics of Fully Ionized Gases 
(Interscience, New York), p.144 

\bibitem[Tsuneta et al.(1991)]{1991SoPh..136...37T} 
Tsuneta, S., et al.\ 1991, \solphys, 136, 37 

\bibitem[Ugarte-Urra et al.(2006)]{2006ApJ...643..1245} 
Ugarte-Urra, I., Winebarger, A.~R., \& Warren, H.~P.\ 2006, \apj, 643, 1245

\bibitem[Vesecky et al.(1979)]{1979ApJ...233..987V} 
Vesecky, J.~F., Antiochos, S.~K., \& Underwood, J.~H.\ 1979, \apj, 233, 987 

\bibitem[Warren et al.(2003)]{2003ApJ...593.1174W} 
Warren, H.~P., Winebarger, A.~R., \& Mariska, J.~T.\ 2003, \apj, 593, 1174 

\bibitem[Winebarger \& Warren(2005)]{2005ApJ...626..543} 
Winebarger, A.~R., \& Warren, H.~P.\ 2005, \apj, 626, 543 

\bibitem[Winebarger et al.(2003)]{2003ApJ...587..439W} 
Winebarger, A.~R., Warren, H.~P., \& Mariska, J.~T.\ 2003, \apj, 587, 439 
  
\end{thebibliography}
\end{document}